\documentclass{article}
\usepackage{amsmath}
\usepackage{amssymb}

\input{tcilatex}

\begin{document}

\title{Sudden Future Singularities}
\author{John D. Barrow \\
DAMTP, Centre for Mathematical Sciences\\
Cambridge University\\
Wilberforce Road\\
Cambridge CB3 0WA\\
UK}
\maketitle

\begin{abstract}
We show that a singularity can occur at a finite future time in an expanding
Friedmann universe even when $\rho >0$ and $\rho +3p>0$. Explicit examples
are constructed and a simple condition is given which can be used to
eliminate behaviour of this sort if it is judged to be unphysical.
\end{abstract}

There have been many recent investigations into the theoretical possibility
that expanding universes can come to a violent end at a finite future time,
without experiencing an expansion maximum or subsequent collapse to an
all-encompassing 'big crunch' singularity. It has been shown that in general
relativity and related metric theories of gravity this looming 'big rip' in
spacetime can be precipitated by the presence of matter fields which violate
the dominant energy condition \cite{rip}. Hence, it can occur when matter
exists with the extreme constitutional property that

\begin{equation}
\rho +p<0,  \label{dom}
\end{equation}%
where $\rho $ is the fluid density and $p$ is the pressure. Such extremal
fluids have been dubbed 'phantom' or 'ghost' fields and their behaviour is
counter intuitive in many respects.

Here, we will show that although the extreme 'phantom' equation of state $%
\rho +p<0$ is sufficient to produce a singularity in the future of a
non-contracting universe, as discussed in refs.\cite{rip}, it is by no means
necessary. A finite-time singularity can arise in the expanding phase of a
Friedmann universe without requiring $\rho +p<0$. In fact, we show that such
a singularity can arise even though the strong-energy condition holds and
the matter in the universe obeys $\rho >0$ and $\rho +3p>0,$ regardless of
the sign of the 3-curvature of the universe.

Consider the Friedmann universe with expansion scale factor $a(t)$,
curvature parameter $k$, and Hubble expansion rate $H=\dot{a}/a$; then ($%
8\pi G=c=1$) the Einstein equations reduce to

\begin{eqnarray}
3H^{2} &=&\rho -\frac{k}{a^{2}},  \label{fried} \\
\dot{\rho}+3H(\rho +p) &=&0,  \label{con} \\
\frac{\ddot{a}}{a} &=&-\frac{(\rho +3p)}{6}.  \label{fr}
\end{eqnarray}

Let us first look informally at whether it is possible for any type of
singularity to develop in which a physical scalar becomes infinite at a
finite future comoving proper time $t_{s},$ when, say, $a(t)\rightarrow
a(t_{s})\neq 0$ or $\infty ,$ and $H(t)\rightarrow H_{s}<\infty $ and $%
H_{s}>0.$ If so, this requires

\begin{equation}
\rho \rightarrow 3H_{s}^{2}+\frac{k}{a_{s}^{2}}=\rho _{s}<\infty
\label{lim1}
\end{equation}%
and

\begin{equation}
\frac{\ddot{a}}{a}\rightarrow -\frac{p}{2}\ -\frac{\rho _{s}}{6}\rightarrow -%
\frac{p}{2}\ -\frac{H_{s}^{2}}{2}-\frac{k}{6a_{s}^{2}}  \label{lim2}
\end{equation}

\begin{equation}
\dot{\rho}\rightarrow -3H_{s}(\rho _{s}+p)  \label{lim3}
\end{equation}

Thus we see that the density is necessarily finite at $t_{s}$ but a pressure
singularity can still occur with

\begin{equation}
p(t)\rightarrow \infty  \label{p}
\end{equation}%
as $t\rightarrow t_{s},$ in accord with (\ref{lim2}). In this case the
pressure singularity is accompanied by infinite acceleration

\begin{equation}
\frac{\ddot{a}}{a}\rightarrow -\infty  \label{acc}
\end{equation}%
and there will be a scalar polynomial (sp) curvature singularity \cite{he}
at $t=t_{s}$ even though the density and the expansion rate remain perfectly
finite there, and the scale factor is finite and non-zero. In particular, we
note from this argument that $\rho _{s}>0$ and

\begin{equation}
\rho _{s}+3p_{s}>0.  \label{sec}
\end{equation}

Guided by these heuristic arguments, we now construct an explicit example by
seeking, over the time interval $0<t<t_{s},$ a solution for the scale factor 
$a(t)$ of the form

\begin{equation}
a(t)=A+Bt^{q}+C(t_{s}-t)^{n},  \label{ex}
\end{equation}%
where $A>0,B>0,q>0,C$ and $n>0$ are free constants to be determined. We fix
the zero of time by requiring $a(0)=0$ so $A=-Ct_{s}^{n}>0$ and so

\begin{equation}
H_{s}=\frac{qBt_{s}^{q-1}}{A+Bt_{s}}.  \label{H}
\end{equation}%
Without loss of generality we can use the scaling freedom remaining in the
Friedmann metric to divide by $A$ and then set $A\equiv 1$ $=-Ct_{s}^{n},$
so we have

\begin{equation}
a(t)=(\frac{t}{t_{s}})^{q}\left( a_{s}-1\right) +1-(1-\frac{t}{t_{s}})^{n}
\label{sol2}
\end{equation}%
where $a_{s}\equiv a(t_{s})$. Hence, as $t\rightarrow t_{s}$ from below, we
have

\begin{equation}
\ddot{a}\rightarrow q(q-1)Bt^{q-2}-\frac{\ n(n-1)}{t_{s}^{2}(1-\frac{t}{t_{s}%
})^{2-n}}\rightarrow -\infty  \label{Lim}
\end{equation}%
whenever $1<n<2$ and $0<q\leq 1$; the solution exists on the interval $%
0<t<t_{s}$. Hence, as $t\rightarrow t_{s\text{ }}$we have $a\rightarrow
a_{s} $; $H_{s}$ and $\rho _{s}>0$ (for $3q^{2}(a_{s}-1)^{2}t_{s}^{-2}>-k$)
are finite but $p_{s}\rightarrow \infty $. When $2<n<3,$ we note that $\ddot{%
a}$ remains finite but $\dddot{a}\rightarrow \infty $ as $t\rightarrow t_{s%
\text{ }}$as $p_{s}$ remains finite but $\dot{p}_{s}\rightarrow \infty $. By
contrast, there is an initial all-encompassing strong-curvature singularity,
with $\rho \rightarrow \infty $ and $p\rightarrow \infty $, as $t\rightarrow
0$. From (\ref{Lim}) and (\ref{fr}), we see that $\rho $ and $\rho +3p$
remain positive. This behaviour can occur even in a closed universe ($k=+1$%
): the pressure singularity halts the expansion before an expansion maximum
is reached.

The solution (\ref{sol2}) expands with

\begin{equation*}
a(t)\approx n\frac{t}{t_{s}}+\left( \frac{t}{t_{s}}\right)
^{q}(a_{s}-1)\approx \left( \frac{t}{t_{s}}\right) ^{q}
\end{equation*}%
as $t\rightarrow 0$ and so with choices $q=1/3,1/2,$ $2/3,$or $q=1$ it
resembles a scalar-, radiation-, dust-, or negative-curvature-dominated
Friedmann universe respectively at early times. At late times, as the
'big-rip' singularity is approached, the expansion approaches a constant with

\begin{equation*}
a(t)\approx a_{s}+q(1-a_{s})(1-\frac{t}{t_{s}})
\end{equation*}%
as $t\rightarrow t_{s}$.

This specific family of solutions shows that it is possible for an expanding
universe to develop a 'big-rip' singularity at a finite future time even if
the matter fields in the universe satisfy the strong-energy conditions $\rho
>0$ and $\rho +3p>0$. We see that this pressure-driven sp-curvature
singularity \cite{he} can arise in universes of any curvature. In
particular, in the case with $q=1$ this situation was shown by Barrow,
Galloway and Tipler \cite{bgt} to provide a counter-example to the belief
that $\rho >0$ and $\rho +3p>0$ were sufficient to ensure that a closed ($%
k=+1$) Friedmann universe collapses to a second singularity \cite{ellis, rob}%
. The source of this peculiar behaviour is the uncontrolled increase of the
pressure which can occur independently of the matter density in the absence
of an appropriate equation of state linking $p$ and $\rho $. There are two
independent Friedmann equations for the three quantities $a,p,$ and $\rho $.
In the absence of a relation between any two of them there is an
unconstrained degree of freedom. If we introduce an equation of state $%
p(\rho )$ which bounds the pressure by some well-behaved function of the
density -- for example $p<C\rho $ with $C>0$ -- then the pressure
singularity at a finite future time is eliminated if we require $\rho >0$
and $\rho +3p>0$. An upper bound on the ratio $p/\rho $ corresponds to a
bound on signal propagation of waves associated with small changes in
pressure. It is interesting to note that situations with $p/\rho >>1$ arise
in the ekpyrotic universe scenario \cite{ek}.

This phenomenon is not especially sensitive to the isotropic nature of the
expansion. If we consider the simplest anisotropic universes with isotropic
3-curvature then the shear evolves as $\sigma \propto a^{-3}$, where $a$ is
now the geometric-mean scale factor. If pressure singularities occur with $a$
finite at a finite $t_{s\text{ \ }}$they will therefore be accompanied by
finite shear stresses and contribute only finite terms to the generalised
Friedmann and acceleration equations. Thus they cannot stop the appearance
of a big-rip singularity \cite{dab}. If we add a traceless anisotropic
pressure tensor, $\pi _{ab}$, then no new effect occurs for the usual
physical models for $\pi _{ab}$ because it is either proportional to the
fluid density $\pi _{ab}=C_{ab}\rho $ and will remain finite when $\rho
<\infty $ (see for example the discussion in ref. \cite{barpi, jmar}) or in
the case of shear viscosity, proportional to the shear scalar. An
anisotropic pressure singularity will only be possible under a generalised
version of the conditions given in this paper. For example, a form $\pi
_{ab}=D_{ab}p$ with no bound on the values of the principal pressures can
diverge close to isotropy when $p\rightarrow \infty $. However, there is the
added constraint of the second law of thermodynamics to be satisfied if
dissipation occurs. No new ingredient is introduced by introducing a bulk
viscous stress $p=(\gamma -1)\rho -3H\eta $, with bulk viscosity $\eta
=\alpha \rho ^{m}$, $\alpha >0$ and $m$ constants. This includes as a
special case the so-called Cardossian \cite{free} or Chaplygin stress \cite%
{chap}, when $\gamma =1$ and $m=-1,$ or $m$ takes an arbitrary negative
value, respectively, in flat Friedmann models \cite{jb} and can avoid future
singularities when $\rho +p<0$, \cite{diaz}. Finite time singularities have
been discussed in the context of braneworld cosmologies, tachyonic
cosmology, and effective 4-d string theory in refs. \cite{calcag, pasq,
kofinas, sahni}; there are other contributions to the field equations in
these theories which allow for their occurrence.

In conclusion, we have shown that, contrary to the impression gained from
the existing literature, \cite{rip}, finite-time future singularities can
appear under the surprisingly weak conditions of $\rho >0$ and $\rho +3p>0$
in an expanding universe. We have identified conditions that are sufficient
to avoid the appearance of this type of sp-curvature singularity at finite
future time. These results can either be used as a more fine-grained method
to search for finite-time singularities and evaluate their likelihood of
occurrence with realistic fluids or, if such singularities are regarded as
causal pathologies, to impose realistic conditions which exclude them by
fiat.

\textit{Acknowledgements} I would like to thank G. Calcagni, D. Coule, M. D%
\c{a}browski, V. Gorini, A. Kamenshchik, G. Kofinas, U. Moschella and V.
Sahni for helpful comments and references.

\end{document}